\documentclass[conference]{IEEEtran}
\IEEEoverridecommandlockouts

\usepackage{pdflscape}
\usepackage{afterpage}
\usepackage{cite}
\usepackage{amsmath,amssymb,amsfonts}
\usepackage{algorithmic}
\usepackage{graphicx}
\usepackage{textcomp}
\usepackage{xcolor}
\usepackage{listings}
\usepackage{xurl,enumitem}
\usepackage[utf8]{inputenc}
\usepackage{float}
\usepackage{booktabs}
\usepackage{pgfplots}
\pgfplotsset{compat=1.18}

\usepackage[normalem]{ulem}

\definecolor{codegreen}{rgb}{0,0.6,0}
\definecolor{codegray}{rgb}{0.5,0.5,0.5}
\definecolor{codepurple}{rgb}{0.58,0,0.82}
\definecolor{backcolour}{rgb}{0.95,0.95,0.92}
\definecolor{codeblue}{rgb}{0.0, 0.0, 0.8}

\lstdefinestyle{mystyle}{
  backgroundcolor=\color{backcolour}, commentstyle=\color{codegreen},
  keywordstyle=\color{magenta},
  numberstyle=\tiny\color{codegray},
  stringstyle=\color{codepurple},
  basicstyle=\ttfamily\footnotesize,
  breakatwhitespace=false,
  breaklines=true,
  captionpos=b,
  keepspaces=true,
  numbers=left,
  numbersep=5pt,
  showspaces=false,
  showstringspaces=false,
  showtabs=false,
  tabsize=2,
  columns=fullflexible
}

\def\BibTeX{{\rm B\kern-.05em{\sc i\kern-.025em b}\kern-.08em
T\kern-.1667em\lower.7ex\hbox{E}\kern-.125emX}}

\renewcommand{\baselinestretch}{0.9}
\begin{document}
\setlength{\baselineskip}{12.05pt}
\title{%Nvidia Sionna Testbed Interaction with an AMD Xilinx FPGA Implementation of OFDM with MU-MIMO Beamforming
An FPGA-in-the-Loop Testbed for MU-MIMO OFDM Beamforming over Ray-Traced Wireless Channels\thanks{This work is supported by National Science Foundation under Grant Numbers CNS-2202972, CNS- 2318726, and CNS-2232048.}
}

\author{
  \IEEEauthorblockN{Drew Schlesener, Tolunay Seyfi, Fatemeh Afghah}
  \IEEEauthorblockA{{Holcombe Department of Electrical and Computer Engineering, 
    Clemson University, Clemson SC USA} \\
  \{schlese, tseyfi, fafghah\}@clemson.edu, 
}
}
\maketitle
%%make stronger argument instead of just summarize what I have done. What I am doing, why is it novel, what is the problem I am trying to solve

\begin{abstract}
  Wireless networks face ever expanding throughput demands from heterogeneous, high-density user populations, requiring beamforming algorithms that adapt to channel conditions with low latency. Validating such algorithms requires either costly over-the-air testbeds or simulation environments that lack the timing and resource constraints of real hardware, leaving a gap between algorithm design and hardware-realizable deployment. This work presents a frame-based hardware-in-the-loop (HIL) testbed that closes that gap by coupling an FPGA-based implementation of OFDM Waveforms with MU-MIMO beamforming to NVIDIA Sionna's ray-tracing channel simulator, enabling a physical base-station architecture to transmit and receive against a Sionna-rendered digital-twin propagation environment. Unlike prior work that validates beamforming algorithms either purely in simulation or on full RF testbeds, this architecture allows beamforming logic running on actual FPGA fabric to be evaluated under realistic, controllable, and repeatable channel conditions, including UE mobility and site-specific multi-path, without requiring an anechoic chamber or live RF front end. We detail the FPGA OFDM transmit/receive pipeline, the synchronization and data interface between the FPGA and the Sionna environment, and validation of signal quality under AWGN and ray-traced channel conditions, establishing this testbed as a platform for hardware-validated beamforming research.
\end{abstract}


%%need 10-15 citations
\section{Introduction}

As wireless networks scale in device density, traffic volume, and service heterogeneity, they face increasing throughput demands under stringent energy and latency constraints. This is particularly important in modern 5G and emerging 6G systems, where flexible waveform processing, beamforming, and adaptive link control must be efficiently executed \cite{LDPC,DORA_25}. Multiple-user multiple-input multiple-output (MU-MIMO) beamforming enables multiple users to share the same time-frequency resources through spatial precoding \cite{9390169}. In OFDM-based systems, this requires complex weighting across antenna branches while preserving symbol timing, subcarrier structure, and modulation accuracy. Hardware realization introduces effects absent from idealized floating-point models, including fixed-point quantization, pipeline latency, finite logic resources, synchronization constraints, and implementation-dependent error floors. These effects make FPGA-based validation particularly relevant for MU-MIMO OFDM, where FFT/iFFT processing, complex weight multiplication, and per-antenna signal generation require deterministic streaming execution \cite{9076114,10864253}.

FPGAs provide hardware-level parallelism, deterministic low-latency execution, and post-deployment reconfigurability \cite{10864253}. Unlike general-purpose CPUs and GPUs, which depend on software scheduling, memory hierarchies, and often batched execution, FPGAs can map low-PHY functions such as FFT/iFFT, channel estimation, beamforming, filtering, coding, and symbol processing directly into customized hardware pipelines. This can reduce data-movement overhead and improve responsiveness for streaming workloads \cite{AtlasRAN}. Although ASICs can provide greater efficiency and optimization, FPGAs offer greater flexibility for prototyping, standards evolution, field updates, and algorithm-hardware co-design \cite{9610039,10323918}. However, hardware implementation alone does not provide a realistic and repeatable wireless channel. Full over-the-air testbeds provide real propagation effects but require RF front ends, antenna arrays, calibration, controlled environments, and costly repeated experiments \cite{electronics8121490,7832644}, while ray-tracing and link-level simulation provide configurable, repeatable channels without exercising the actual fixed-point FPGA MU-MIMO OFDM data-path \cite{11417943}. This work bridges these approaches by coupling FPGA-based MU-MIMO OFDM beamforming with NVIDIA Sionna ray-traced channels in a hardware-in-the-loop (HIL) workflow, providing an accessible platform for hardware-realistic beamforming evaluation under controllable and repeatable propagation conditions without replacing a full RF testbed.

The HIL testbed uses a Digilent Arty A7-100T FPGA as a physical base-station platform interfaced over USB JTAG with a host PC running NVIDIA Sionna's ray-tracing channel simulator. The platform was selected for its balance of reconfigurability, hardware accessibility, and suitability for FPGA-based Low-PHY SDR experimentation. The main contributions are:
\begin{itemize}[leftmargin=*]
    \item \textbf{Digital-Twin Sionna Channel Modeling:} Sionna parses the FPGA's OFDM MIMO output and uses its Monte-Carlo ray-tracing \texttt{PathSolver()} to evaluate the hardware-generated signals in a realistic digital-twin propagation environment.
    \item \textbf{Accessible FPGA HIL Testbed:} The architecture couples FPGA beamforming hardware with controllable, repeatable, site-specific Sionna channels, including UE mobility, without requiring an anechoic chamber or live RF front end.
    \item \textbf{Hardware Performance and Signal-Integrity Validation:} The FPGA is benchmarked against an equivalent CPU baseline, achieving a 13.1$\times$ throughput improvement and 571$\times$ lower energy per symbol with post-route timing closure at 100~MHz. BER/EVM are characterized for QPSK, 16-QAM, and 64-QAM across four SNR levels, up to four concurrent UEs, and two Sionna-rendered propagation scenes.
\end{itemize}
\section{Related Work}

\subsection{Hardware Acceleration for Low-PHY Functions}

FPGAs are highly effective for latency-critical low physical-layer (Low-PHY) wireless functions, such as OFDM signal processing, due to their high-throughput parallelism and reconfigurability. For instance, FPGA-based OpenCL kernels significantly outperform sequential processors in time-sensitive OpenAirInterface (OAI) gNB processing \cite{11026993}, while matching GPU performance with superior energy efficiency \cite{shah2023heterogeneous}. Because FPGAs excel in deterministic, streaming computation with low power demands, this research builds upon these advantages to develop an FPGA-centric platform supporting both Low-PHY operations and beamforming. 

\subsection{Modern FPGA Frameworks and RFSoC}

Meeting the stringent architectural demands of massive MIMO systems increasingly relies on advanced FPGA and RFSoC frameworks. Recent work illustrates a shift toward unified hardware capable of end-to-end wireless functionality, such as 3GPP-compliant integrated 5G base stations \cite{10.1145/3737900.3770173}. Furthermore, modern implementations are moving beyond conventional signal processing by incorporating hardware-aware optimizations like fixed-point arithmetic to reduce MIMO transceiver complexity and power consumption \cite{11085972}. Building on this progression, the present work targets a scalable, integrated FPGA-based OFDM and beamforming platform.

\section{Methodology}

\subsection{Fixed-Point MU-MIMO OFDM Data-path}

The FPGA-oriented base-band design was developed in MATLAB Simulink 2024b using Vitis Model Composer and synthesized in Vivado 2025.1. The data-path implements a configurable multi-stream OFDM transmitter with antenna-domain spatial precoding for a $4\times4$ planar array. The implemented design supports quadrature phase-shift keying (QPSK), 16-quadrature amplitude modulation (16-QAM), and 64-QAM. Each active UE is assigned an independent deterministic input payload during validation, allowing the transmitted bit sequence associated with each spatial stream to be reproduced exactly at the receiver for per-UE bit-error-rate measurements.

Let $M \in \{4,16,64\}$ denote the modulation order and let $q=\log_2(M)$ be the number of bits per complex symbol. For each active UE stream, groups of $q$ input bits are first mapped to an unnormalized Gray-coded constellation point
\begin{equation}
x_M^\star = a_I + j a_Q,
\end{equation}
where $a_I,a_Q \in \mathcal{A}_M$ are the unnormalized in-phase and quadrature amplitude levels. The corresponding hardware symbol is obtained by applying a modulation-dependent scaling factor,
\begin{equation}
x_M = K_M^{\mathrm{HW}} x_M^\star
      = K_M^{\mathrm{HW}}(a_I+j a_Q),
\end{equation}
where $K_M^{\mathrm{HW}}$ denotes the scaling factor implemented in the FPGA datapath.

For square QAM constellations, the amplitude alphabet is
$\mathcal{A}_M = \{\pm 1,\pm 3,\ldots,\pm(\sqrt{M}-1)\}$.
For QPSK, viewed as square 4-QAM, this reduces to
$\mathcal{A}_4=\{\pm1\}$. The scaling factor is selected to normalize the average symbol energy while satisfying the fixed-point dynamic range of the AMD Xilinx FFT intellectual property core. The unnormalized average symbol energy is
\begin{equation}
E^\star_{s,M}
=
\mathbb{E}\!\left[|a_I+j a_Q|^2\right].
\end{equation}
This gives $E^{\star}_{s,4}=2$, $E^{\star}_{s,16}=10$, and
$E^{\star}_{s,64}=42$ for the supported constellations.

The corresponding theoretical unit-energy normalization factor is
\begin{equation}
K_M^{\mathrm{th}}=\frac{1}{\sqrt{E^{\star}_{s,M}}}.
\end{equation}

This yields the theoretical values
\[
K_4^{\mathrm{th}}=\frac{1}{\sqrt{2}},\qquad
K_{16}^{\mathrm{th}}=\frac{1}{\sqrt{10}},\qquad
K_{64}^{\mathrm{th}}=\frac{1}{\sqrt{42}}.
\]

For QPSK and 16-QAM, the implemented hardware scaling matches the theoretical normalization, i.e.,
$K_4^{\mathrm{HW}}=K_4^{\mathrm{th}}$ and
$K_{16}^{\mathrm{HW}}=K_{16}^{\mathrm{th}}$.
For the implemented 64-QAM datapath, a slightly smaller hardware scaling value of
$K_{64}^{\mathrm{HW}}=0.13$
is used instead of the theoretical value
$K_{64}^{\mathrm{th}}=0.1543$
to provide additional guard margin against bit growth and overflow in the fixed-point IFFT and spatial-precoding stages.

The modulated symbols are represented using the Vitis fixed-point
$\mathrm{Fix}_{16,15}$ format, corresponding to a signed 16-bit
word with 15 fractional bits. After constellation mapping and scaling, the active subcarrier symbols associated with each UE are passed through the OFDM modulation stage to produce independent time-domain waveforms. For $N_{\mathrm{UE}}$ simultaneously active users, these waveforms are represented by the spatial-stream vector
\begin{equation}
\mathbf{s}[n]
=
\begin{bmatrix}
s_0[n] &
s_1[n] &
\cdots &
s_{N_{\mathrm{UE}}-1}[n]
\end{bmatrix}^{T},
\end{equation}
where $s_u[n]$ denotes the time-domain OFDM sample associated with UE $u$.

The active spatial streams are jointly mapped to the sixteen antenna branches through a complex multi-user precoding matrix
\begin{equation}
\mathbf{W}
=
\begin{bmatrix}
\mathbf{w}_0 &
\mathbf{w}_1 &
\cdots &
\mathbf{w}_{N_{\mathrm{UE}}-1}
\end{bmatrix}
\in
\mathbb{C}^{16\times N_{\mathrm{UE}}},
\end{equation}
where $\mathbf{w}_u \in \mathbb{C}^{16}$ denotes the spatial precoding vector associated with UE $u$. The resulting antenna-domain transmit vector is
\begin{equation}
\mathbf{x}[n]
=
\mathbf{W}\mathbf{s}[n]
\in
\mathbb{C}^{16}.
\end{equation}
Equivalently, the signal generated for antenna branch $p$ is
\begin{equation}
x_p[n]
=
\sum_{u=0}^{N_{\mathrm{UE}}-1}
w_{p,u}s_u[n],
\qquad
p=0,\ldots,15.
\end{equation}

The precoding coefficients are represented in fixed-point form and applied within the FPGA datapath, allowing the hardware-generated antenna-domain signals to incorporate the quantization, finite-word-length, and pipeline effects of the implemented spatial-precoding stage. The multi-user precoding coefficients are updated by the beamforming-control stage according to the active UE and channel configuration.

At the receiver side of the validation model, the hardware-generated antenna-domain signals are propagated through the corresponding channel realization and processed for each active UE. The received waveform for UE $u$ is transformed back to the frequency domain using an FFT and de-mapped according to the selected modulation order. The recovered bits are compared with the deterministic payload associated with UE $u$ to compute per-UE bit error rate, while the recovered constellation points are used to compute per-UE error vector magnitude. This modular design allows the fixed-point OFDM and multi-user spatial-precoding pipeline to be evaluated across different modulation orders, active UE counts, precoding parameters, and channel conditions.

\subsection{FPGA Implementation and Host Interface}

\begin{figure*}[t]
  \centering
  \includegraphics[width=0.65\textwidth]{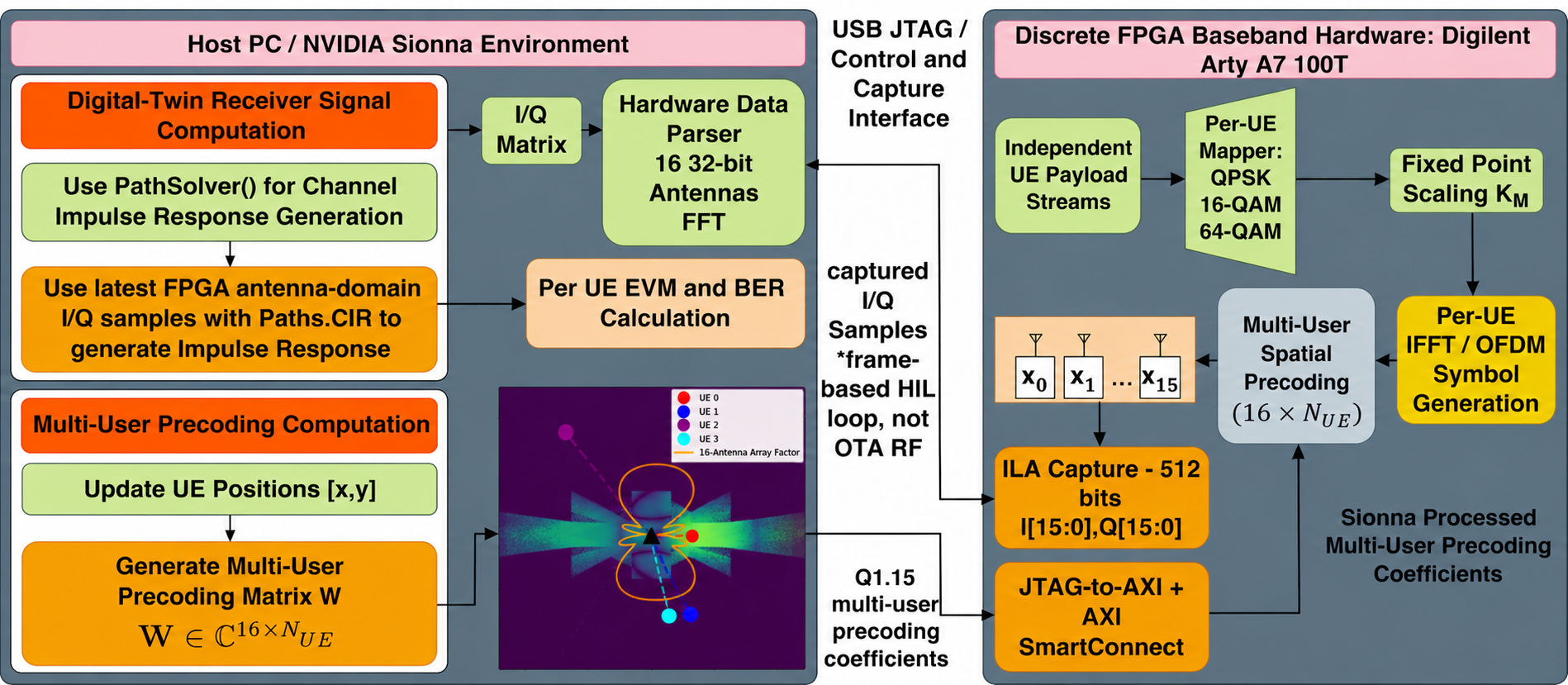}
  \caption{FPGA Hardware-in-Loop Block Diagram.}
  \label{fig:FPGAHIL}
  \vspace{-0.4cm}
\end{figure*}

After verification in the Simulink/Vitis Model Composer environment, the fixed-point OFDM and multi-user spatial-precoding data-path was exported as synthesized hardware and integrated into a Vivado block design targeting the Digilent Arty A7-100T FPGA. The top-level hardware design includes the exported Vitis Model Composer intellectual property block, a clocking wizard, reset logic, a JTAG-to-AXI Master, an AXI interconnect, and an Integrated Logic Analyzer (ILA) for internal signal capture. The design is driven by a 100~MHz clock generated from the board reference clock, corresponding to a 10~ns clock period for the implemented base-band pipeline.

The multi-user precoding coefficients are exposed to the host through an AXI-mapped control interface. For each hardware-in-the-loop frame, the host-side Sionna workflow uses the current UE and channel configuration to generate the complex spatial-precoding coefficients for the active UE streams. The coefficients are quantized to the Q1.15 fixed-point format and written to the FPGA through the JTAG-to-AXI Master using Vivado TCL commands. Inside the FPGA, the coefficients are applied by the fixed-point multi-user spatial-precoding data-path to jointly map the active OFDM streams to the sixteen antenna branches of the 4$\times$4 transmit array.

The resulting antenna-domain in-phase/quadrature samples are routed to the ILA for internal signal capture. The captured samples are exported to the host and used as the hardware-generated transmit-array input to the Sionna ray-traced channel model. Thus, for each HIL frame, the host updates the channel-dependent precoding parameters, the FPGA performs the corresponding fixed-point multi-user spatial-precoding operation, and the resulting sixteen antenna-domain signals are returned to the host for propagation through the ray-traced channel and subsequent signal-quality evaluation. The interface is intended for controlled frame-based hardware validation rather than real-time radio operation. Because precoding-parameter updates and sample extraction are performed through JTAG, TCL scripting, and ILA capture, the loop does not represent a standards-compliant 5G New Radio front-haul or over-the-air implementation. Instead, it provides an accessible mechanism for validating the FPGA OFDM and multi-user spatial-precoding data-path under repeatable channel conditions before transitioning to higher-throughput interfaces, radio-frequency front ends, or radio-frequency system-on-chip platforms.
\subsection{Timing Closure}
 In Vivado, the Post-route static timing analysis confirms that all user-specified timing constraints 
are met for the implemented design. The primary design clock operates at 100~MHz 
(10~ns period), generated via an Mixed-Mode Clock Manager (MMCM) clocking primitive from the board's 100~MHz reference 
oscillator. Table~\ref{tab:timing} summarizes the worst-case setup and hold margins.
\vspace{-0.2cm}
\begin{table}[h]
\centering
\caption{Post-Route Timing Summary (Worst Negative Slack)}
\label{tab:timing}
\begin{tabular}{lcc}
\hline
\textbf{Check} & \textbf{Slack (ns)} & \textbf{Status} \\
\hline
Setup (WNS) & 0.199 & Met \\
Hold (WHS)  & 0.016 & Met \\
Pulse Width (WPWS) & 3.000 & Met \\
\hline
\end{tabular}
\vspace{-0.18cm}
\end{table}

 The critical setup path traverses 65 logic levels, dominated by a 64-stage 
\texttt{CARRY4} ripple-carry chain consistent with a wide fixed-point accumulation 
structure in the     multi-user spatial-precoding accumulation or FFT butterfly data-path. With only 
0.199~ns of setup slack at 100~MHz, this carry chain represents the limiting factor for 
any future increase in operating frequency; pipelining this accumulator with intermediate 
register stages is the primary candidate for improving maximum achievable throughput. 
Hold margin is similarly tight at 0.016~ns, indicating limited robustness to placement 
perturbation or re-implementation and warranting attention in any future design iteration.

\subsection{FPGA Design Interfacing with Sionna}

The final stage of the framework interfaces the deployed FPGA design with NVIDIA Sionna to evaluate the hardware-generated MU-MIMO OFDM signals under ray-traced propagation conditions. For each hardware-in-the-loop frame, the Sionna workflow updates the active UE configuration and corresponding channel information. Channel-dependent multi-user precoding coefficients are generated by the beamforming-control stage, quantized to the hardware fixed-point representation, and written to the FPGA through the AXI interface using Vivado TCL commands. The FPGA applies these coefficients to the active UE streams and generates the corresponding sixteen antenna-domain I/Q signals. The latest hardware-generated I/Q samples are captured by the Vivado ILA and exported to the host for propagation through the Sionna channel model, as illustrated in Fig.~\ref{fig:FPGAHIL}.

Episodes are defined as sequences of $N$ frames, with the number of active UEs and SNR varied across the experimental test matrix. Each frame is updated every 1000~ms. During a frame update, the UE positions are advanced, the corresponding channel and precoding parameters are updated, the latest FPGA-generated antenna-domain samples are applied to the channel simulation environment, and the resulting signal-quality metrics are logged. For $N_{\mathrm{UE}}$ active users, the FPGA transmit signal is represented by the antenna-domain matrix
\begin{equation}
\mathbf{X}
=
\begin{bmatrix}
\mathbf{x}[0]^T & \mathbf{x}[1]^T & \cdots & \mathbf{x}[T-1]^T
\end{bmatrix}^{T}
\in \mathbb{C}^{T\times N_{\mathrm{ant}}}.
\end{equation}
where $N_{\mathrm{ant}}=16$ and each row contains the jointly precoded antenna-domain samples generated from the active UE streams.

Sionna's \texttt{PathSolver} ray-traces the scene deterministically to produce a channel impulse response (CIR) for each UE. After the scene is populated with the digital-twin transmitter, active UEs, and physical geometry, the solver computes the corresponding propagation paths. Each propagation path is characterized by a complex channel coefficient $a_i$, propagation delay $\tau_i$, and associated angles of departure and arrival. The resulting path set is represented as
\begin{equation}
\text{Paths}
=
\left\{
(a_i,\tau_i,\theta_{T,i},\varphi_{T,i},
\theta_{R,i},\varphi_{R,i})
\right\}_{i=1}^{N_{\mathrm{paths}}}.
\end{equation}

Because the antenna array is explicitly modeled, the solver determines propagation paths associated with the transmit antenna elements for each UE, yielding a per-antenna channel response. The resulting path coefficients are extracted using \texttt{paths.cir()} and combined over the path and delay dimensions to form the channel vector for UE $u$,
\begin{equation}
\mathbf{h}_u
=
\sum_{\ell,\tau}
a_{u,\ell,\tau}
\in
\mathbb{C}^{N_{\mathrm{ant}}}.
\end{equation}

The channel vectors of the simultaneously active UEs may be collected into the multi-user channel matrix
\begin{equation}
\mathbf{H}
=
\begin{bmatrix}
\mathbf{h}_0^{T} & \mathbf{h}_1^{T} & \cdots & \mathbf{h}_{N_{\mathrm{UE}}-1}^{T}
\end{bmatrix}^{T}
\in
\mathbb{C}^{N_{\mathrm{UE}}\times N_{\mathrm{ant}}}.
\end{equation}

For UE $u$, the FPGA-generated antenna-domain signals are propagated through the corresponding ray-traced channel to form the received signal
\begin{equation}
\mathbf{y}_{u,\mathrm{clean}}
=
\mathbf{X}\mathbf{h}_u.
\end{equation}
Since $\mathbf{X}$ contains the jointly precoded contributions of all simultaneously active UE streams, $\mathbf{y}_{u,\mathrm{clean}}$ represents the received waveform associated with the complete multi-user transmission at UE $u$.

Additive White Gaussian Noise (AWGN) is injected according to the SNR specified for the current episode,
\begin{equation}
\mathbf{y}_{u,\mathrm{noisy}}
=
\mathbf{y}_{u,\mathrm{clean}}
+
\mathbf{n}_u.
\end{equation}
The received waveform is then transformed to the frequency domain using an FFT, and the active subcarriers $\mathcal{S}_{\mathrm{active}}$ are extracted,
\begin{equation}
Y_u[k]
=
\sum_{n=0}^{N_{\mathrm{FFT}}-1}
y_u[n]e^{-j2\pi kn/N_{\mathrm{FFT}}},
\qquad
k\in\mathcal{S}_{\mathrm{active}}.
\end{equation}

The active subcarrier vectors are normalized and evaluated with respect to the constellation and payload associated with the intended stream of each UE. Let $d_{u,k}$ denote the reference constellation symbol transmitted for UE $u$ on active subcarrier $k$, and let $b_{u,i}$ and $\hat{b}_{u,i}$ denote the corresponding transmitted and detected bits, respectively. The per-UE Error Vector Magnitude (EVM) is computed as
\begin{equation}
EVM_{u,\mathrm{dB}}
=
10\log_{10}
\left(
\frac{1}{N}
\sum_{k=1}^{N}
\left|
\widetilde{Y}_{u,k}-d_{u,k}
\right|^2
\right),
\end{equation}
and the per-UE Bit Error Rate (BER) is
\begin{equation}
BER_u
=
\frac{1}{N_{b,u}}
\sum_{i=1}^{N_{b,u}}
\mathbf{1}
\left[
\hat{b}_{u,i}\neq b_{u,i}
\right].
\end{equation}

These metrics are evaluated separately for each active UE during simultaneous multi-user transmission. The Matplotlib visualization is updated each frame with UE positions, the selected UE's constellation and received waveform, rolling EVM, and per-UE EVM/BER. The title reports the episode index, active UE count, and SNR.

\section{Evaluation}
 The experimental platform was hosted on a Dell Precision 3650 Tower equipped with an Intel i9-10900 CPU, an NVIDIA Quadro P6000 GPU, and 16\,GB of RAM, running Ubuntu 24.04 LTS. The OFDM hardware subsystem was developed in MATLAB Simulink 2024b using Vitis Model Composer, while the FPGA block design was implemented in Vivado 2025.1. The final hardware implementation targeted the discrete Digilent Arty A7-100T FPGA platform.
 
 System performance was evaluated using FPGA cycle count for MU-MIMO OFDM symbol completion. The Sionna channel uses a 3.5~GHz carrier with 30~kHz subcarrier spacing. FPGA and CPU symbol-completion times are compared in nanoseconds, and FPGA EVM is compared with EVM from OFDM generation entirely within Sionna. The FPGA is evaluated over 16 episodes of 50 frames, repeated six times for QPSK, 16-QAM, and 64-QAM in two scenes: a custom empty plane with a wall and \texttt{simple\_street\_canyon}. The test is divided into four 4-episode SNR blocks, each separated by 10~dB. Within each block, the number of simultaneously active UE streams increases from 1--4, and BER/EVM are evaluated versus SNR and UE count.

\section{Experimental Results}

\subsection{Performance and Benchmarking Results}

\vspace{-0.18cm}
\begin{table}[ht]
    \centering
    \caption{CPU vs. FPGA Performance Metrics for 16-Point CP-OFDM}
    \label{tab:performance_metrics}
    \resizebox{\columnwidth}{!}{%
    \renewcommand{\arraystretch}{1.3}
    \begin{tabular}{@{}lllc@{}}
        \toprule
        \textbf{Metric} & \textbf{CPU (Intel i9-10900)} & \textbf{FPGA (Artix-7 Core)} & \textbf{Advantage} \\
        \midrule
        \textbf{Clock Frequency} & 2.8 GHz (up to 5.2 GHz) & 100 MHz & CPU ($\sim$52$\times$) \\
        \textbf{Processing Type} & Sequential (Time-Domain) & Unrolled Parallel & FPGA \\
        \textbf{Throughput (16-Ant)} & 131.82 ns & \textbf{10.00 ns} & \textbf{FPGA (13.1$\times$)} \\
        \textbf{Pipeline Latency} & N/A (Jittered) & 1,090.00 ns & FPGA \\
        \textbf{Operating Power} & $\sim$65 W (TDP) & $\sim$1.5 W & FPGA ($\sim$43.3$\times$) \\
        \textbf{Energy/Symbol} & 8.56 $\mu$J & \textbf{0.015 $\mu$J} & \textbf{FPGA (571$\times$)} \\
        \bottomrule
    \end{tabular}%
    }
\end{table}

The FPGA outputs a new 16-antenna MU-MIMO OFDM symbol each 100~MHz clock cycle, with a 109-cycle pipeline latency of $1.09~\mu\text{s}$. CPU OFDM generation was timed per symbol in C++ using the \texttt{chrono} library. The equivalent software implementation processes independent UE OFDM streams and jointly maps them to the 16 antenna outputs through multi-user spatial precoding. Assuming zero overhead for the vectorized 16-antenna multiplication gives a best-case CPU bottleneck at the IFFT boundary of $T_{\text{CPU}}\ge131.82$~ns, excluding OS interrupts, cache misses, scheduling, and bus latency. As shown in Table~\ref{tab:performance_metrics}, the FPGA achieves a deterministic $13.1\times$ speedup through fully parallel MU-MIMO OFDM processing. While SIMD accelerates software beamforming, CPUs remain constrained by sequential iFFT execution and software-managed data movement, whereas the FPGA uses unrolled spatial-processing datapaths for predictable latency and higher throughput.

\begin{figure}[ht]
    \centering
    
    % First Graph: Throughput (Left)
    \begin{minipage}[t]{0.45\columnwidth}
        \centering
        \resizebox{\linewidth}{!}{% <--- ADDED RESIZEBOX
        \begin{tikzpicture}[baseline=(current axis.north)]
            \begin{axis}[
                ybar,
                ymode=log,
                enlargelimits=0.3,
                legend style={at={(0.5,-0.32)}, anchor=north, legend columns=-1, font=\tiny},
                ylabel={Execution Time (ns) [Log Scale]},
                symbolic x coords={Throughput},
                xtick=data,
                nodes near coords,
                nodes near coords align={vertical},
                title={Throughput Comparison},
                bar width=14pt, 
                width=\linewidth,
                height=4.0cm 
                ]
            \addplot coordinates {(Throughput, 131.82)};
            \addplot coordinates {(Throughput, 10.00)};
            \legend{Intel i9-10900, Artix-7 FPGA}
            \end{axis}
        \end{tikzpicture}%
        } % <--- CLOSE RESIZEBOX
    \end{minipage}
    \hfill 
    % Second Graph: Energy (Right)
    \begin{minipage}[t]{0.45\columnwidth}
        \centering
        \resizebox{\linewidth}{!}{% <--- ADDED RESIZEBOX
        \begin{tikzpicture}[baseline=(current axis.north)]
            \begin{axis}[
                ybar,
                ymode=log,
                enlargelimits=0.3,
                legend style={at={(0.5,-0.32)}, anchor=north, legend columns=-1, font=\tiny},
                ylabel={Energy ($\mu$J) [Log Scale]},
                symbolic x coords={Energy/Symbol},
                xtick=data,
                nodes near coords,
                nodes near coords align={vertical},
                title={Energy Efficiency},
                bar width=14pt, 
                width=\linewidth,
                height=4.0cm 
                ]
            \addplot coordinates {(Energy/Symbol, 8.56)};
            \addplot coordinates {(Energy/Symbol, 0.015)};
            \legend{Intel i9-10900, Artix-7 FPGA}
            \end{axis}
        \end{tikzpicture}%
        } % <--- CLOSE RESIZEBOX
    \end{minipage}
    
    \vspace{0.4cm}
    \caption{Logarithmic scale comparison of overall block throughput time (left) and computational energy consumed per block (right). Lower values represent superior performance.}
    \label{fig:performance_graphs}
    \vspace{-0.3cm} 
\end{figure}
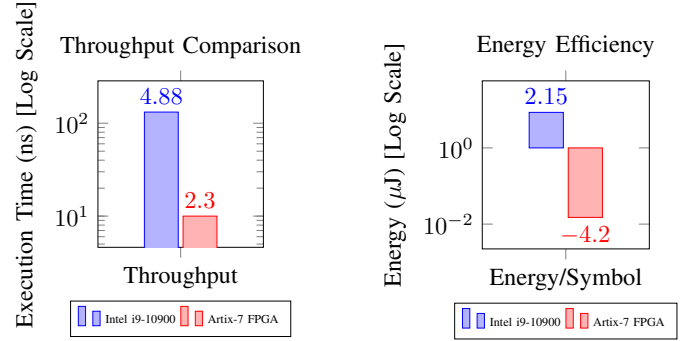

To compare metrics spanning several orders of magnitude, Figure~\ref{fig:performance_graphs} reports execution time and energy per symbol on a base-10 logarithmic scale, preserving quantitative contrast between the CPU and FPGA implementations without compressing the smaller values. Even under optimistic best-case CPU assumptions, the FPGA implementation displays an advantage in both throughput and energy efficiency.

\subsection{Data Rate and Interface Bandwidth}

The base-band generator uses a 16-point IFFT ($N_{\text{active}}=16$ active subcarriers); with QPSK ($Q_m=2$ bits/subcarrier), each spatial stream carries $B_{\text{sym}} = N_{\text{active}} \times Q_m = 32$ bits/symbol, giving a maximum information rate $R_{\text{max}} = B_{\text{sym}}/T_{\text{proc}}$. For the best-case CPU ($T_{\text{proc}}=131.82\text{ ns}$), $R_{\text{CPU}} \approx \mathbf{242.7\text{ Mbps}}$; the FPGA's fully unrolled pipeline, producing a symbol every $10\text{ ns}$, reaches $R_{\text{FPGA}} = \mathbf{3.2\text{ Gbps}}$. Scaled to $4$ UEs via independent precoding, CPU capacity fragments to $\sim$$60.6\text{ Mbps}$/user, while the FPGA's parallel user pipelines scale cleanly to $\mathbf{12.8\text{ Gbps}}$ aggregate, validating its necessity for dense MU-MIMO deployments.

This payload-level rate excludes the raw I/Q volume moved internally: the beamforming stage drives 16 antennas, each fed a 32-bit complex sample (16-bit I, 16-bit Q), for $512$ bits of I/Q data per symbol. At one symbol per $10\text{ ns}$, this yields a physical interface bandwidth of $\mathbf{51.2\text{ Gbps}}$. Sustaining this rate through a general-purpose processor would require repeated movement through software-managed memory, cache, and I/O, introducing scheduling overhead and jitter; the FPGA instead keeps this bandwidth local to the fabric, generating all sixteen antenna branches in parallel with deterministic clock-cycle timing, illustrating its architectural advantage for MU-MIMO OFDM beamforming without host-side data movement per symbol.

\subsection{Episodic Testing}

\begin{figure}[ht]
  \centering
  \includegraphics[width=0.8\columnwidth]{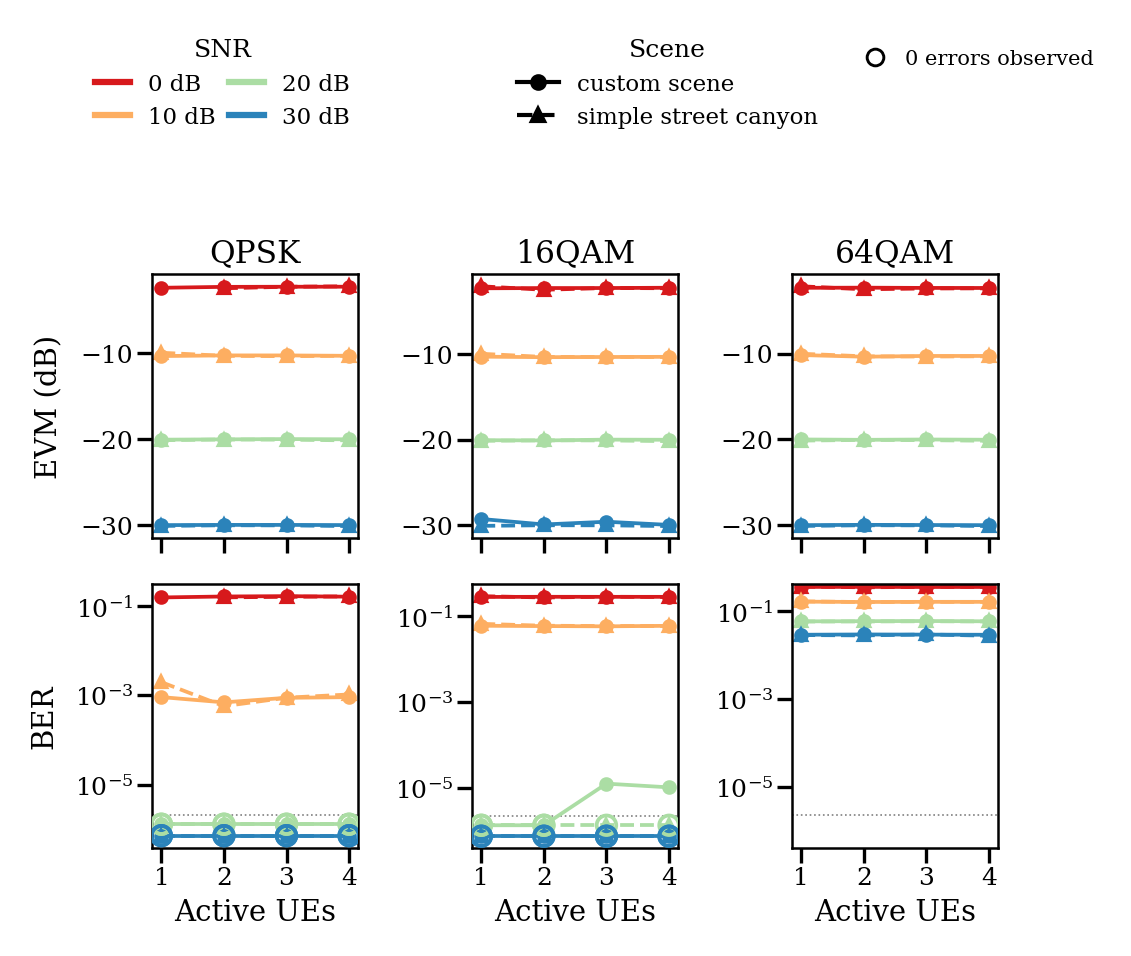}
  \caption{FPGA Sionna EVM Comparisons with different SNR and UE count}
  \label{fig:FPGAEpisodic}
\end{figure}

As the testing script progresses through each four-episode SNR block, Figure~\ref{fig:FPGAEpisodic} shows the expected reduction in EVM as SNR increases. Within the current measurement pipeline, BER and EVM are evaluated separately for each simultaneously transmitted UE stream using the corresponding channel realization. Under this evaluation methodology, the measured BER and EVM exhibit limited dependence on the number of active UEs over the tested range of one to four users. BER also shows limited variation across the two evaluated scene geometries.

\begin{table}[h]
\centering
\caption{Mean BER (with $\pm$1 std.\ across the two test scenes, $n=2$) by Modulation and active UE count (UE1-UE4), faceted by SNR. Each cell's mantissa is scaled by the row's listed power of ten. Each entry is the mean of the per-episode.}
\label{tab:ber_summary}
\scriptsize
\setlength{\tabcolsep}{2pt}
\begin{tabular}{c c c c c c c}
\toprule
SNR & Modulation & $\times$ & U1 & U2 & U3 & U4 \\
\midrule
0 & QPSK & $10^{-1}$ & $1.5\textsuperscript{a}$ & $1.6{\scriptstyle\pm}<0.1$ & $1.6{\scriptstyle\pm}<0.1$ & $1.6{\scriptstyle\pm}<0.1$ \\
 & 16Q & $10^{-1}$ & $2.9{\scriptstyle\pm}0.1$ & $2.8{\scriptstyle\pm}0.1$ & $2.9{\scriptstyle\pm}<0.1$ & $2.8{\scriptstyle\pm}<0.1$ \\
 & 64Q & $10^{-1}$ & $3.6{\scriptstyle\pm}<0.1$ & $3.6{\scriptstyle\pm}0.1$ & $3.6{\scriptstyle\pm}<0.1$ & $3.6{\scriptstyle\pm}<0.1$ \\
\midrule
10 & QPSK & $10^{-3}$ & $1.5{\scriptstyle\pm}0.8$ & $0.64{\scriptstyle\pm}0.1$ & $0.89{\scriptstyle\pm}<0.1$ & $0.99{\scriptstyle\pm}0.1$ \\
 & 16Q & $10^{-2}$ & $6.4{\scriptstyle\pm}0.5$ & $6{\scriptstyle\pm}0.1$ & $5.9{\scriptstyle\pm}0.1$ & $6{\scriptstyle\pm}<0.1$ \\
 & 64Q & $10^{-1}$ & $1.7{\scriptstyle\pm}<0.1$ & $1.6{\scriptstyle\pm}<0.1$ & $1.6{\scriptstyle\pm}<0.1$ & $1.6{\scriptstyle\pm}<0.1$ \\
\midrule
20 & QPSK & -- & $0$ & $0$ & $0$ & $0$ \\
 & 16Q & $10^{-6}$ & $0$ & $0$ & $6.3{\scriptstyle\pm}8.9$ & $5.2{\scriptstyle\pm}7.3$ \\
 & 64Q & $10^{-2}$ & $5.9{\scriptstyle\pm}<0.1$ & $5.9{\scriptstyle\pm}0.1$ & $6{\scriptstyle\pm}0.1$ & $5.9{\scriptstyle\pm}<0.1$ \\
\midrule
30 & QPSK & -- & $0$ & $0$ & $0$ & $0$ \\
 & 16Q & -- & $0$ & $0$ & $0$ & $0$ \\
 & 64Q & $10^{-2}$ & $2.9{\scriptstyle\pm}<0.1$ & $2.9{\scriptstyle\pm}0.1$ & $2.9{\scriptstyle\pm}<0.1$ & $2.8{\scriptstyle\pm}0.1$ \\
\bottomrule
\end{tabular}
\vspace{-0.3cm}
\end{table}

 Modulation order has little effect on EVM but strongly affects BER because denser constellations have tighter decision regions. As shown in Table~\ref{tab:ber_summary}, BER decreases monotonically with SNR for all modulations: QPSK falls from 0.15--0.16 at 0~dB to 0 by 20~dB, while 16-QAM falls from 0.28--0.29 to a near-zero floor, with UE1--UE2 reaching 0 and UE3--UE4 showing isolated errors on the order of $10^{-6}$. 64-QAM decreases more slowly and remains at 0.029 at 30~dB, where QPSK and 16-QAM are effectively error-free. This SNR-independent residual suggests a hardware limitation beyond additive noise, such as residual coarse-to-fine phase-lock error, I/Q quantization, or other RF impairments, warranting further characterization. Accordingly, higher-order modulation requires higher SNR for equivalent BER, with QPSK $<$ 16-QAM $<$ 64-QAM at every tested SNR; at 10~dB, QPSK is near $10^{-3}$ while 64-QAM remains at 0.16--0.17.

\section{Conclusion}

This paper presents a hardware-in-the-loop testbed coupling FPGA-based MU-MIMO OFDM beamforming with NVIDIA Sionna's ray-traced channels, enabling physical FPGA beamforming logic to be evaluated in a controllable, repeatable digital-twin environment without an anechoic chamber or live RF front end. Characterization across QPSK, 16-QAM, and 64-QAM confirmed the expected SNR- and modulation-dependent BER behavior, validating the FPGA OFDM transmit/receive and demodulation pipeline. BER and EVM showed no significant dependence on active UE count, indicating that the measurement pipeline scales cleanly with additional users. A persistent, SNR-independent 64-QAM BER floor, absent for QPSK and 16-QAM, indicates a hardware-level limitation such as residual phase-lock error or I/Q quantization requiring further characterization. These results establish the testbed as a platform for hardware-realistic beamforming research. Future work will investigate higher-speed feedback interfaces, including PCIe and Ethernet, and optimized beamforming control. 

\bibliographystyle{ieeetr}
\bibliography{references}

@INPROCEEDINGS{7832644,
  author={Thompson, Daniel and Yeary, Mark and Fulton, Caleb},
  booktitle={2016 IEEE International Symposium on Phased Array Systems and Technology (PAST)}, 
  title={RF array system equalization and true time delay with FPGA hardware-in-the-loop}, 
  year={2016},
  volume={},
  number={},
  pages={1-5},
  doi={10.1109/ARRAY.2016.7832644}}

@ARTICLE{11417943,
  author={Ryu, Sehyun and Yang, Hyun Jong},
  journal={IEEE Transactions on Vehicular Technology}, 
  title={Standards-Compliant DM-RS Allocation via Temporal Channel Prediction for Massive MIMO Systems}, 
  year={2026},
  volume={},
  number={},
  pages={1-5},
  doi={10.1109/TVT.2026.3654149}}

@Article{electronics8121490,
AUTHOR = {Duarte, Luis and Gomes, Rodolfo and Ribeiro, Carlos and Caldeirinha, Rafael F. S.},
TITLE = {A Software-Defined Radio for Future Wireless Communication Systems at 60 GHz},
JOURNAL = {Electronics},
VOLUME = {8},
YEAR = {2019},
NUMBER = {12},
ARTICLE-NUMBER = {1490},
URL = {https://www.mdpi.com/2079-9292/8/12/1490},
ISSN = {2079-9292},
DOI = {10.3390/electronics8121490}
}

@INPROCEEDINGS{10323918,
  author={Lin, Chunxiao and Azmine, Muhammad Farhan and Yi, Yang},
  booktitle={2023 IEEE/ACM International Conference on Computer Aided Design (ICCAD)}, 
  title={Invited Paper: Accelerating Next-G Wireless Communications with FPGA-Based AI Accelerators}, 
  year={2023},
  volume={},
  number={},
  pages={1-8},
  doi={10.1109/ICCAD57390.2023.10323918}}

@INPROCEEDINGS{10864253,
  author={Vaithianathan, Muthukumaran and Udkar, Shivakumar and Roy, Deepanjan and Reddy, Manjunath and Rajasekaran, Senkadir},
  booktitle={2024 International Conference on Sustainable Communication Networks and Application (ICSCNA)}, 
  title={FPGA Prototyping of DSP Algorithms for Wireless Communication Systems}, 
  year={2024},
  volume={},
  number={},
  pages={231-236},
  doi={10.1109/ICSCNA63714.2024.10864253}}

@ARTICLE{9076114,
  author={Chamola, Vinay and Patra, Sambit and Kumar, Neeraj and Guizani, Mohsen},
  journal={IEEE Wireless Communications}, 
  title={FPGA for 5G: Re-configurable Hardware for Next Generation Communication}, 
  year={2020},
  volume={27},
  number={3},
  pages={140-147},
  doi={10.1109/MWC.001.1900359}}

@INPROCEEDINGS{11026993,
  author={Bhattacharyya, Abhishek and Fumagalli, Andrea and Kondepu, Koteswararao},
  booktitle={2025 IEEE 26th International Symposium on a World of Wireless, Mobile and Multimedia Networks (WoWMoM)}, 
  title={DEMO: FPGA-accelerated 5G Low-PHY Functions and An Integration with OpenAirInterface}, 
  year={2025},
  volume={},
  number={},
  pages={160-162},
  doi={10.1109/WoWMoM65615.2025.00038}}

@INPROCEEDINGS{11085972,
  author={Pathi, A M V and Prasad, D.Durga and Mukhesh, M.Sai and Nandan, M.Utkarsh and Kumar, M.Dileep and Subrahmanyam, K M V S},
  booktitle={2025 6th International Conference on Intelligent Communication Technologies and Virtual Mobile Networks (ICICV)}, 
  title={Efficient VLSI Architecture for OTFS-MIMO Transceiver in Multipath Fading Channels}, 
  year={2025},
  volume={},
  number={},
  pages={1338-1343},
  doi={10.1109/ICICV64824.2025.11085972}}

@inproceedings{10.1145/3737900.3770173,
author = {Praneeth, K. Sai and Sattianarayanin, Jeeva Keshav and Kotturi, Sai Prasanth and Singh, Rohit and Gundlapalle, Aravind and Singh, Siddharth and Ganti, Radha Krishna},
title = {Design of a 32 Channel 5G NR MIMO Base Station},
year = {2025},
isbn = {9798400719776},
publisher = {Association for Computing Machinery},
address = {New York, NY, USA},
url = {https://doi.org/10.1145/3737900.3770173},
doi = {10.1145/3737900.3770173},
booktitle = {Proceedings of the 2nd ACM Workshop on Open and AI RAN},
pages = {43–49},
numpages = {7},
location = {
},
series = {OpenRan '25}
}

@phdthesis{shah2023heterogeneous,
  title={Heterogeneous Acceleration for 5G New Radio Channel Modelling Using FPGAs and GPUs},
  author={Shah, Nasir Ali},
  year={2023},
  school={Politecnico di Torino}
}

@ARTICLE{9610039,
  author={Papatheofanous, Elissaios Alexios and Reisis, Dionysios and Nikitopoulos, Konstantinos},
  journal={IEEE Access}, 
  title={LDPC Hardware Acceleration in 5G Open Radio Access Network Platforms}, 
  year={2021},
  volume={9},
  number={},
  pages={152960-152971},
  doi={10.1109/ACCESS.2021.3127039}}

@misc{DORA_25,
      title={{{DORA}}: Dynamic O-RAN Resource Allocation for Multi-Slice 5G Networks}, 
      author={Alireza Ebrahimi Dorcheh and Tolunay Seyfi and Fatemeh Afghah},
      year={2025},
      eprint={2509.07242},
      archivePrefix={arXiv},
      primaryClass={cs.NI},
      url={https://arxiv.org/abs/2509.07242}, 
}

@misc{LDPC,
      title={Six Times to Spare: LDPC Acceleration on DGX Spark for AI-Native Open RAN}, 
      author={Ryan Barker and Fatemeh Afghah},
      year={2026},
      eprint={2602.04652},
      archivePrefix={arXiv},
      primaryClass={cs.DC},
      url={https://arxiv.org/abs/2602.04652}, 
}

@misc{AtlasRAN,
      title={AtlasRAN: Modeling and Performance Evaluation of Open 5G Platforms for Ubiquitous Wireless Networks}, 
      author={Ryan Barker and Tolunay Seyfi and Alireza Ebrahimi Dorcheh and Julia Boone and Fatemeh Afghah and Joseph Boccuzzi},
      year={2026},
      eprint={2603.14661},
      archivePrefix={arXiv},
      primaryClass={cs.NI},
      url={https://arxiv.org/abs/2603.14661}, 
}

@ARTICLE{9390169,
  author={Tataria, Harsh and Shafi, Mansoor and Molisch, Andreas F. and Dohler, Mischa and Sjöland, Henrik and Tufvesson, Fredrik},
  journal={Proceedings of the IEEE}, 
  title={6G Wireless Systems: Vision, Requirements, Challenges, Insights, and Opportunities}, 
  year={2021},
  volume={109},
  number={7},
  pages={1166-1199},
  doi={10.1109/JPROC.2021.3061701}}

\end{document}